\documentclass[manuscript]{aastex631}

\submitjournal{RNAAS}
\accepted{March 9, 2025}

\begin{document}

\title{2005 VL1 is not Venera-2}


\author[0000-0002-7093-295X]{Jonathan McDowell}
\affiliation{Smithsonian Astrophysical Observatory}
\email{planet4589@gmail.com}

\begin{abstract}

The solar system object 2005 VL1 passed close to Earth in late 1965. It has been suggested
that it is actually the space probe Venera-2. However, a comparison of the orbits presented
in this note demonstrates that the proposed association is incorrect.

\end{abstract}


\section{Venera-2 and 2005 VL1}

\cite{LC25} 
argue that the object 2005 VL1 is the lost Soviet
space probe Venera-2. However, the orbital data do not support this idea, as pointed out by
\citep{Deen2025} and others.
In particular, \cite{LC25} dismiss the reported
4 degree ecliptic inclination of Venera-2 as a likely measurement error, when in fact
it was critical in ensuring a close approach to Venus in early 1966.

In this note I investigate what is known about Venera-2's actual trajectory.

The USSR's 3MV-4 No. 4 space probe, built by the Korolev OKB-1 organization, was named Venera-2 in
public sources. Venera-2 was launched on 1965 Nov 12 at 0446:48 UTC by an 8K78 rocket from
the Baykonur spaceport to a 51 degree low Earth orbit \citep{GCAT}. The 8K78 fourth stage,
designated the Blok-L, fired at about 0607 UTC the same day to inject
itself and Venera-2 on hyperbolic Earth escape trajectories.
Tracking of Venera-2 by NORAD indicates that the burn would have been from a 250 km altitude
circular orbit with 51.9 deg inclination, and would have occurred
over the equator at 8 deg W 0 deg N (geocentric RA 133 deg).

Soviet press releases indicate that Venera-2's closest approach to
Venus was about 24000 km above the cloudtops on 1966 Feb 27 at 0252 UTC.
It is not clear from Soviet sources whether any
significant trajectory corrections were made during the Earth-Venus cruise. Communications
with the probe were spotty near encounter and lost soon afterwards \citep{Siddiqi02}.

\section{Reconstruction of the trajectory of Venera-2}

At departure (1965 Nov 12 at 0607 UTC) Earth was at Cartesian heliocentric ecliptic coordinates
of E = (0.634, 0.759, 0.000) AU
and at arrival (1966 Feb 27 0252 UTC) Venus
was at Cartesian heliocentric ecliptic coordinates of
V = (-0.718, 0.025, 0.042) AU.

I am not aware of a Soviet source for the Earth-to-Venus transfer orbit of Venera-2,
but Baker (1977, Spaceflight v 17, p 446) cites an orbit of 0.7183 x 1.0190 AU x 4.12 deg.
I will refer to this as the `Venera-2 quoted orbit'.
Assuming these parameters are osculating elements near the time of launch,
I use JPL Horizons to propagate the orbit
with suitable choices of node, argument of perihelion and mean anomaly
such that it passes close to (within a Hill radius of) points E and V at approximately the correct times.
This gives classical elements (semi major axis, eccentricity, inclination, node, argument of perihelion
and mean anomaly)
\begin{quote}
$a=0.8687$ au, $e=0.1731$, $i=4.12^o$, $\Omega = 50.1^o$, $\omega=162.3^o$, $M=205.2^o$.
\end{quote}

I will call this choice of
parameters the `Venera-2 model orbit'. This is not an optimized fit but suffices to  
give a general idea of the actual trajectory. The elements are relative to the J2000 ecliptic and have epoch 1965 Nov 12.

The orbital parameters of 2005 VL1, propagated to 1965 Nov by JPL Horizons, are very different:
\begin{quote}
$a=0.8994$ au, $e=0.2238$, $i=0.29^o$, $\Omega = 1.2^o$, $\omega=311.6^o$, $M=124.1^o$.
\end{quote}

and 2005 VL1 does not pass near Venus in early 1966. In particular, its low
ecliptic inclination means that it cannot be a 1965-66 Venus probe.

The figure shows the trajectories projected onto the ecliptic X,Y and X,Z planes.
Note the use of the 4 degree inclination trajectory to arc up out of the plane
and down to Venus, which is at a 3 degree ecliptic latitude at encounter. 

\begin{figure*}[ht!]
\plotone{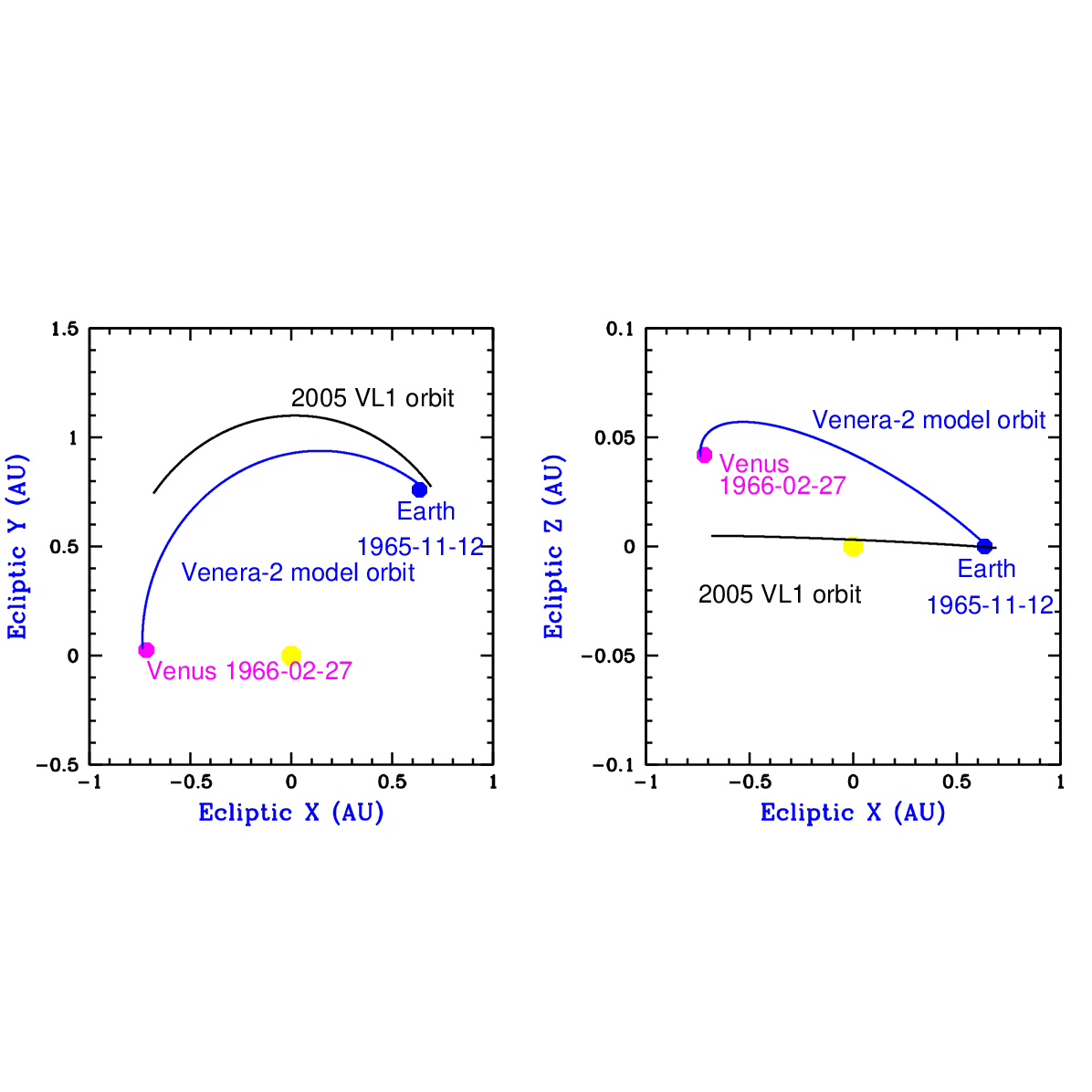}
\caption{Venera-2 and 2005 VL1 trajectory from 1965 Nov 12 to 1966 Feb 27, projected on the ecliptic XY plane (left) and XZ plane (right).
Note that the Z axis is exaggerated by a factor 10 compared to the X and Y axes.
}
\end{figure*}

\section{Additional lost objects in similar orbits}

Finally, I note that there are two other objects expected to be in 
similar orbits to Venera-2 \citep{GCAT}. The Blok-L rocket final stage from
Venera-2's launch vehicle is of a similar size and should be in a
similar orbit to Venera-2 itself. Venera-2's sister probe Venera-3 was
launched on 1965 Nov 16 and, after a trajectory correction, impacted
Venus on 1966 Mar 1. Venera-3's Blok-L final stage rocket will again be
in a similar orbit to Venera-3's pre-trajectory-correction orbit, which
was reported to have a Venus miss distance of 65000 km. Both these
Blok-L stages will share the Venera-2 probe's ecliptic inclination, and
so neither of them can be 2005 VL1.

\section{Discussion}

Although this particular association between a lost space probe and an observed solar system object is spurious,
searching for such cases is reasonable.
The author's General Catalog of Artificial Space Objects \citep{GCAT} includes 345 known artificial objects currently
in heliocentric orbit (and a further 10 on hyperbolic solar system escape trajectories). The majority of these
objects have not been tracked in many years, and their current orbits are poorly known. There are a number
of documented cases of candidate minor planets that turned out instead to be members of this lost
artificial object population, for example J002E3 \citep{Jorgensen03} and 2018 CN41 \citep{MPEC38,MPEC49}.
Traffic in cislunar and interplanetary space is rapidly increasing with many new actors, governmental and commercial,
often without public release of the associated trajectories,
and hence the rate of these misidentfications is also likely to increase. 
As advocated  by the American Astronomical Society in a recent statement \citep{AAS24},
sharing deep space probe trajectories
publicly will be important to avoid contaminating the statistics of the near earth object population.

\software{Horizons \citep{Horizons,Giorgini15}, SM \citep{SM}     }


\bibliography{ms}{}
\bibliographystyle{aasjournal}

\end{document}